\documentclass[preprint,showpacs,preprintnumbers,amsmath,amssymb]{revtex4}
\usepackage{graphicx}% Include figure files
\usepackage{floatflt}
\usepackage{dcolumn}% Align table columns on decimal point
\usepackage{bm}% bold math
\usepackage{rotating}
\usepackage{natbib}
\usepackage{txfonts}
\begin{document}
   \title{Correlations in the properties of static and 
rapidly rotating compact stars }

   \author{B. K. Agrawal}
              \email{bijay.agrawal@saha.ac.in}
   \affiliation{Saha Institute of Nuclear Physics, Kolkata - 700064, India.}
   \author{ Raj Kumar and Shashi K. Dhiman}
             \email{shashi.dhiman@gmail.com}
     \affiliation{ Department of Physics, Himachal Pradesh University, Shimla - 171005, India.}

\date{\today}

  \begin{abstract}

  Correlations in the properties of the static compact stars (CSs)
and the ones rotating with the highest observed frequency of 1122Hz
are studied using a large set of equations of state (EOSs).  These EOSs
span various approaches  and their
chemical composition vary from the nucleons to hyperons and quarks in
$\beta$-equilibrium. 
It is found that the properties of static CS, like, the maximum gravitational
mass $M_{\rm max}^{\rm stat}$ and radius $R_{1.4}^{\rm stat}$ corresponding to the canonical mass and supramassive
or non-supramassive nature of the CS rotating at 1122 Hz are strongly
correlated. In particular, 
only those EOSs yield the CS rotating at
1122Hz to be non-supramassive for which $\left (\frac{M_{\rm max}^{\rm
stat}}{M_\odot}\right )^{1/2} \left (\frac{10{\rm km }}{R_{1.4}^{\rm
stat}} \right)^{3/2} $ is greater than unity.  Suitable parametric form
which can be used to split the $M_{\rm max}^{\rm stat}$ $-$ $R_{1.4}^{\rm
stat}$  plane into the regions of different supramassive nature   of
the CS rotating at 1122Hz is presented. 
Currently measured maximum gravitational mass
1.76$M_\odot$ of PSR J0437-4715 suggests that the CS rotating at 1122Hz can
be 
non-supramassive provided
$R_{1.4}^{\rm stat} \leqslant 12.4$ km.

\end{abstract}

%   \keywords{ dense matter -- equation of state --
%stars: compact stars : rotation}
\maketitle

\section{Introduction}
\label{intro-sec}
The accurate knowledge of the properties of  static and rotating compact
stars (CSs) are of utmost importance to probe the behaviour of the equation
of state (EOS) of superdense matter.  Even the accurate information on
the maximum gravitational mass $M_{\rm max}^{\rm stat}$ for the static CS
and its radius $R_{1.4}^{\rm stat} $  with the canonical
mass ($1.4M_\odot$), not yet well known,  would narrow down the choices for
the plausible EOSs to just a few. 
The newly measured CS mass 1.76$\pm0.20M_\odot$  of PSR J0437-4715 \cite{Verbiest08}
is obtained by the precise
determination of the orbital inclination angle, the highest measured mass for
any known pulsar to date.
Recent observations of the thermal emission from the quiescent LMXB
X7 in the globular cluster 47 Tuc yield the value of R$_{1.4}$ to be 
14.5$^{+1.8}_{-1.6}$ km \cite{Heinke06}. 
 The binary pulsars  PSR
J0737-3039A,B with masses of the individual star being $1.338M_\odot$
and $1.249M_\odot$ are plausible candidates for the measurement the moment
of inertia due to the spin-orbit coupling effects \cite{Lyne04}. It is
expected that a reasonably accurate value of $R_{1.4}^{\rm stat}$ can
be deduced from the moment of inertia measurement of PSR J0737-3039A
\cite{Morrison04}.

Rotating CSs discovered until recent past  have periods of rotation close
to one millisecond.  The first millisecond pulsar PSR B1937+214 rotating
at the  frequency $\nu=641 Hz$ was discovered in 1982 \cite{Backer82}.
In year 2006, a more rapid pulsar PSR J1748-2446ad rotating at $\nu =716$
Hz was detected \cite{Hessels06}. Such frequencies are too low to
affect significantly the structure of  CSs with $ M  >  1M_\odot$
\cite{Shapiro83},  since, these CSs have the Keplerian (mass-shedding)
frequencies larger than 1000Hz.  Very recent discovery of X-ray transient
XTE J1739-285 by Kaaret et al \cite{Kaaret06} suggests that it contains a
CS rotating at 1122Hz.  Following this discovery,  the structure of the CS
rotating with 1122Hz are studied  using several EOSs \cite{Bejger07}. It
is found that, for some of the EOSs, this CS is supramassive, i.e.,
 \begin{equation}
\delta M_{\rm B}=M_{\rm B,max}^{\rm stat} - M_{\rm B,min}, \label{eq:dmb}
 \end{equation} is less than $0$. In the above equation, $M_{\rm
B,max}^{\rm stat }$ is the maximum baryonic mass of the static CS and
$M_{\rm B,min}$ is the minimum mass for the CS rotating with 1122Hz for
a given EOS.

In the present work we search for the possible correlations in some key
properties of static CSs and the ones rotating with the highest observed
frequency of 1122Hz. The properties of these CSs are computed for a
large set of EOSs, which are constructed using variety of approaches
with the chemical compositions ranging from nucleons to hyperons and
quarks in $\beta$-equilibrium. Our results suggest that the values of
$M_{\rm max}^{\rm stat}$, $R_{1.4}^{\rm stat}$,  and   $\delta M_{\rm
B}$  are strongly correlated. Though, $M_{\rm max}^{\rm stat}$ and
$R_{1.4}^{\rm stat} $ alone does not show any systematic correlations
\cite{Bejger05}.  We also unmask the underlying feature
of the EOS responsible for the supramassive or non-supramassive nature
of the CS rotating with 1122Hz.  ~ ~

\section{Equations of State}
\label{eos}

In this work we consider 24 different EOSs with $M_{\rm max}^{\rm
stat}\geqslant 1.6M_\odot$ which exceeds the recent mass measurements
suggesting only 5$\%$ probability that the mass of pulsar PSR
J1516+02B is below 1.59 M$_\odot$ \cite{Freire07}. These EOSs are
constructed using various approaches which can broadly be grouped
into  (i) models based on variational approach, (ii) relativistic or
non-relativistic mean-field models and (iii) Dirac-Brueckner-Hartree-Fock
model.  The first group  contains EOSs involving  neutrons, protons,
electrons and muons.
 The EOSs considered in this group are:
 BJ-C \cite{Bethe74},
FPS \cite{Lorenz93}, 
BBB2 \cite{Baldo97},
AU, WS, and UU \cite{Wiringa88}, and
APR \cite{Akmal98}.
For the second group we consider the EOSs: 
O \cite{Bowers75}, 
GN3\cite{Glendenning85},
GM1 \cite{Glendenning91}, 
TM1 \cite{Sumiyoshi95},
G2\cite{Furnstahl97}, 
BalbN1H1 \cite{balberg97}, 
GMU110\cite{Pons00},
DH \cite{Douchin01}, 
SSK and GSK1 \cite{Agrawal06}, 
UY, U0 and L0 \cite{Dhiman07}, 
GM1-H,   
 UQM52, and CFL52.  
The EOS GM1-H is composed of nucleons and hyperons in $\beta$ equilibrium.
The nucleon-meson interaction parameters are taken from the GM1 parameter
set whereas hyperon meson couplings are obtained from SU(6) model.
The EOS UQM52 involves noninteracting unpaired quark matter, composed
of massive u, d, and s quarks, is based upon the MIT Bag Model of
quarks. This EOS has been calculated by using model parameters,
bag constant, B = 52 MeV/fm$^{3}$, masses of three quarks, m$_u$ =
m$_d$ = 5.00 MeV/c$^2$, m$_s$ = 150 MeV/c$^2$, and QCD coupling constant
$\alpha_c = 0.1$.  The Color-Flavor-Locked quark matter equation of state
(CFL52) is based upon the free energy as described by \cite{Alford01}
and \cite{Rajagopal01}, by using pairing gap parameter $\Delta$ = 100
MeV. The other model parameters such as  bag constant, quark masses,
and QCD constant are same as used for UQM52 EOS.  In the third group we
consider only one EOS: DBHF by Krastev et al. \cite{Krastev06}.

\section{Results and Discussions}
\label{results}
\begin{figure}[ht]
   \centering
\resizebox{7.5in}{!}{ \includegraphics[]{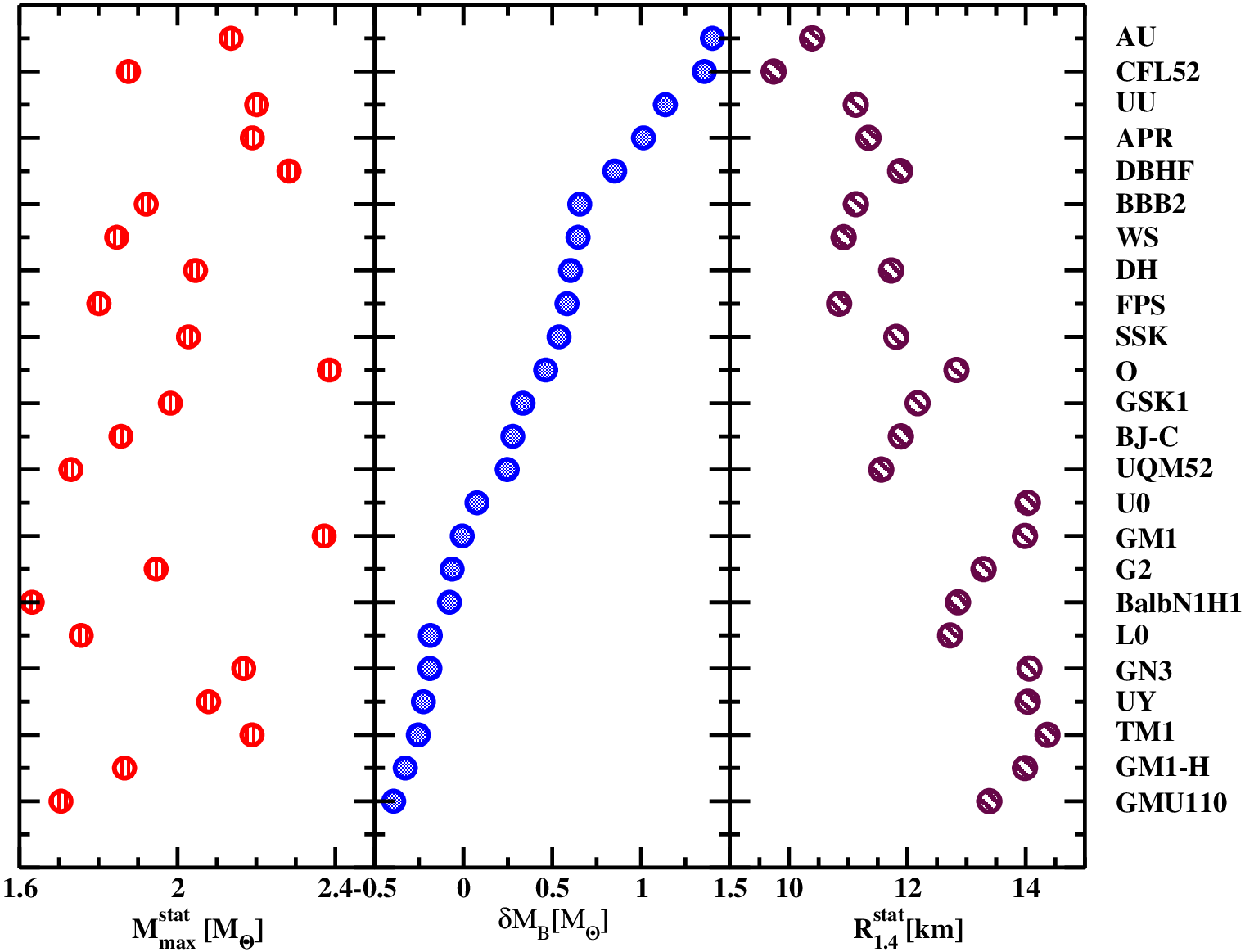}}
  \caption{\label{fig:full_data} (Color online)
Values of the maximum gravitational mass $M_{\rm max}^{\rm stat }$
for static CSs, radius $R_{1.4}^{\rm stat}$ for static CSs with mass
$1.4M_\odot$ and the difference $\delta M_{\rm B}$ (Eq. \ref{eq:dmb})
obtained for several EOSs.}
   \end{figure}

The properties of spherically symmetric static and axially symmetric
rotating CSs are obtained by solving the Einstein's equations in 1D and
2D, respectively. The numerical computations are performed by using RNS
code written by Stergioulas and Friedman \cite{Stergioulas95}.  In Fig.\
\ref{fig:full_data} we present the values of $M_{\rm max}^{\rm stat}$,
$R_{1.4}^{\rm stat}$ and $\delta M_{\rm B}$ obtained for several EOSs.
We notice that the value of $\delta M_{\rm B}$ for a given EOS is only
weakly correlated with those of $M_{\rm max}^{\rm stat}$ as compared to
$R_{1.4}^{\rm stat}$.  It seems, larger  is the value of $R_{1.4}^{\rm
stat}$ smaller will be the $\delta M_{\rm B}$. On the other
hand, $\delta M_{\rm B}$ large means large $M_{\rm max}^{\rm stat}$
but $R_{1.4}^{\rm stat}$ small.  For instance, $\delta M_{\rm B} <
0$ for all those EOSs for which $R_{1.4}^{\rm stat} \gtrsim 14$km.
Whereas, $\delta M_{\rm B} \approx 0.75M_\odot$ for the cases with
$M_{\rm max}^{\rm stat} > 2M_\odot$ and $R_{1.4}^{\rm stat} \sim
10-11$km. Though $\delta M_{\rm B}$ is not correlated well  with $
M_{\rm max}^{\rm stat}$ or $R_{1.4}^{\rm stat}$ alone, but, it might
be  well  correlated  with some appropriate combination of $M_{\rm
max}^{\rm stat}$ and $R_{1.4}^{\rm stat}$.  In Table \ref{tab:pro_1122}
we summarize the properties of the CS, rotating with 1122Hz, calculated
at the maximum circumferential equatorial radius $R_{\rm eq}^{\rm max}$
and the minimum circumferential equatorial radius $R_{\rm eq}^{\rm min}$
for a few selected EOSs. The values of the radius $R_{\rm eq}^{\rm max}$
are determined by the mass shedding instability and that of  $R_{\rm
eq}^{\rm min}$ are determined by the secular axi-symmetric instability
according to  turning point theorem \cite{Friedman88}.  It can be verified
from the  table that the variations in the gravitational mass,
 \begin{equation} \delta M = \left|M(R_{\rm eq}^{\rm max})-M(R_{\rm
 eq}^{\rm min}) \right|, \label{eq:dm} \end {equation} of the CS rotating
 at 1122Hz are
correlated with  $\delta M_{\rm B}$ up to some extent.  The difference
$\delta M \lesssim 0.1 M_\odot$ when $\delta M_{\rm B}$ is negative. 
For $\delta M_{\rm B} > 0$, $\delta M$ increases with $\delta
M_{\rm B}$.  Therefore, $\delta M_{\rm B}$ not only determines whether
the CS rotating at 1122Hz is supramassive or not, but, it also  gives
an estimate about the value of $ \delta M $ for a given EOS.

\begin{figure}[ht]
 \centering
\resizebox{7.5in}{!}{\includegraphics[]{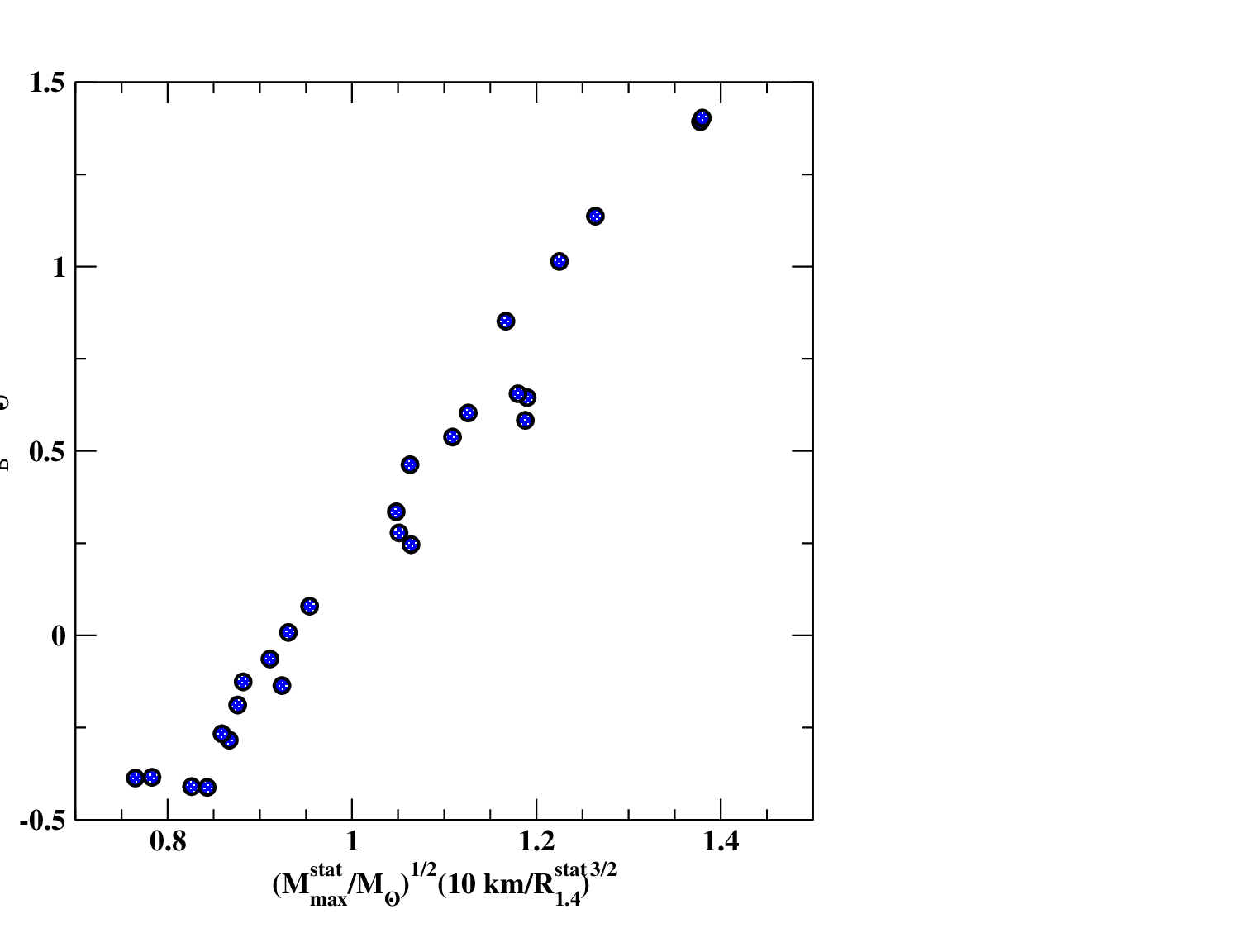}}
\caption{\label{fig:dmb_mr_kap} (Color online)
Correlations between $\delta M_{\rm B}$ and $\left (\frac{M_{\rm max}^{\rm
stat } }{M_\odot}\right )^{1/2} \left (\frac{10{\rm km}}{R_{1.4}^{\rm
stat}}\right )^{3/2}$. This combination of $M_{\rm max}^{\rm stat}$ and
$R_{1.4}^{\rm stat}$ is analogous to the one commonly used to determine
the Keplerian frequency.}
   \end{figure}

In Fig. \ref{fig:dmb_mr_kap} we consider the variations of $\delta M_{\rm
B}$ with $\left (\frac{M_{\rm max}^{\rm stat}}{M_\odot}\right )^{1/2}
\left (\frac{10{\rm km}}{R_{1.4}^{stat}}\right )^{3/2}$.  This combination
of $M_{\rm max}^{\rm stat}$ and $R_{1.4}^{\rm stat}$ is analogous to
the one derived within the Newtonian approximation  to determine the
value of the  Keplerian frequency.  The values of $\delta M_{\rm B}$
and $\left (\frac{M_{\rm max}^{\rm stat }}{M_\odot}\right )^{1/2}
\left (\frac{10{\rm km}}{R_{1.4}^{\rm stat}}\right )^{3/2}$ are well
correlated. It is interesting to note that $\delta M_{\rm B} > 0$
only if   $\left (\frac{M_{\rm max}^{\rm stat}}{M_\odot}\right )^{1/2} \left
(\frac{10{\rm km}}{R_{1.4}^{\rm stat}}\right )^{3/2}$ is greater than
unity.  These correlations simply suggest that the high density behaviour
of a EOS with respect to its behaviour at low density plays a predominant
role in determining whether the CS rotating at 1122Hz is supramassive
or not.  Since, the  $M_{\rm max}^{\rm{stat}}$ probes densest segment
of the EOS whereas, $R_{1.4}^{\rm stat}$ probes relatively lower density
region of EOS.

\begin{figure}[ht]
%   \centering
\resizebox{7.5in}{!}{\includegraphics[]{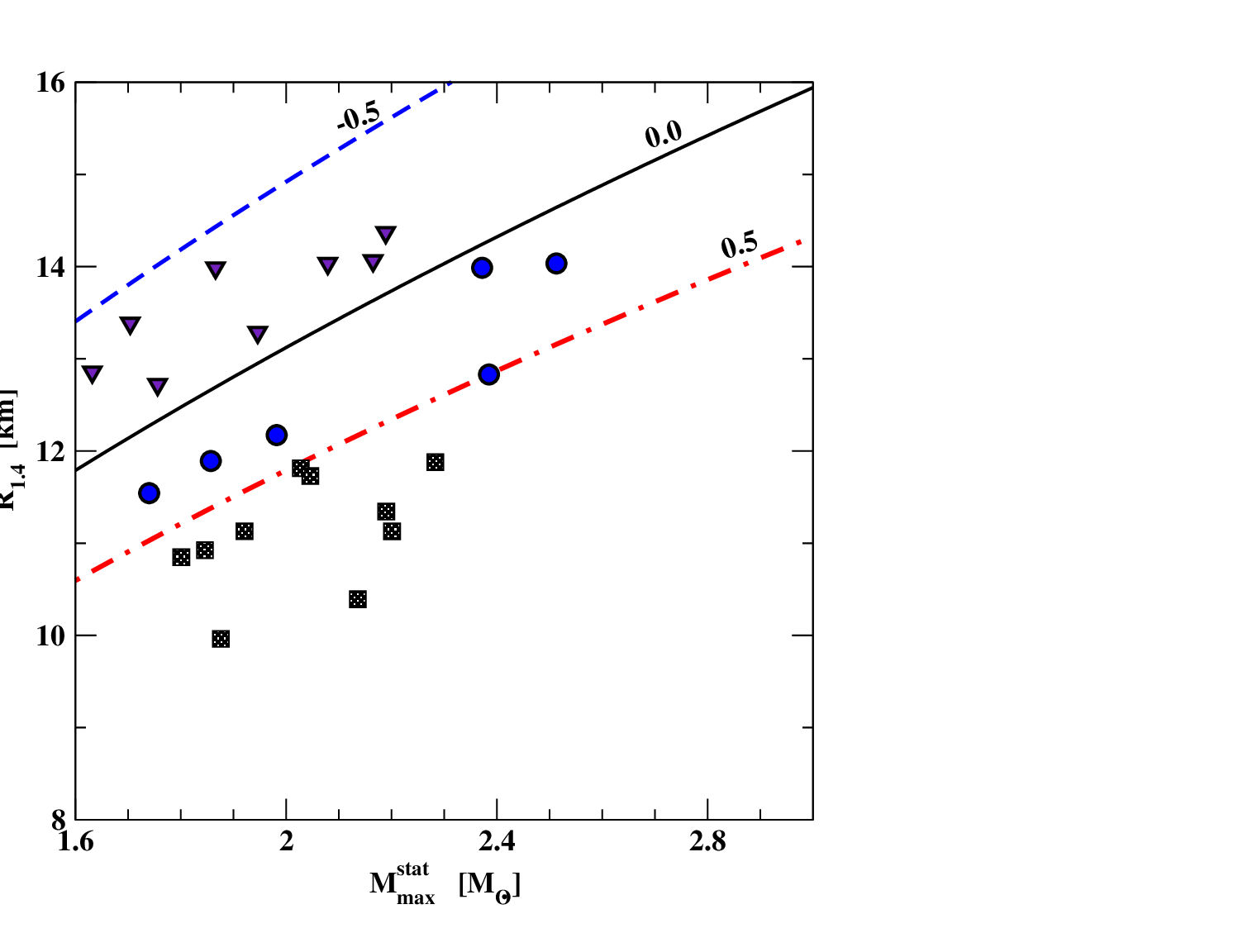}}
\caption{\label{fig:mmax_r14} (Color online)
Plots for $R_{1.4}^{\rm stat}$ versus $M_{\rm max}^{\rm stat}$ 
generated using Eq. \ref{eq:corr} for  $\delta M_{\rm B} = -0.5, 0.0$ and
0.5 $M_\odot$ as indicated.  Different  symbols represent the values of
$M_{\rm max}^{\rm stat}$ and $R_{1.4}^{\rm stat}$ with $\delta M_{B}$ lying
in the range of $-0.5 - 0.0$ (triangles), $0.0 - 0.5$ (circles) and $0.5 -
1.5$ (squares) $M_\odot$  as also depicted in Fig.  \ref{fig:full_data}}
   \end{figure}

\begin{figure}[ht]
   \centering
\resizebox{7.5in}{!}{\includegraphics[]{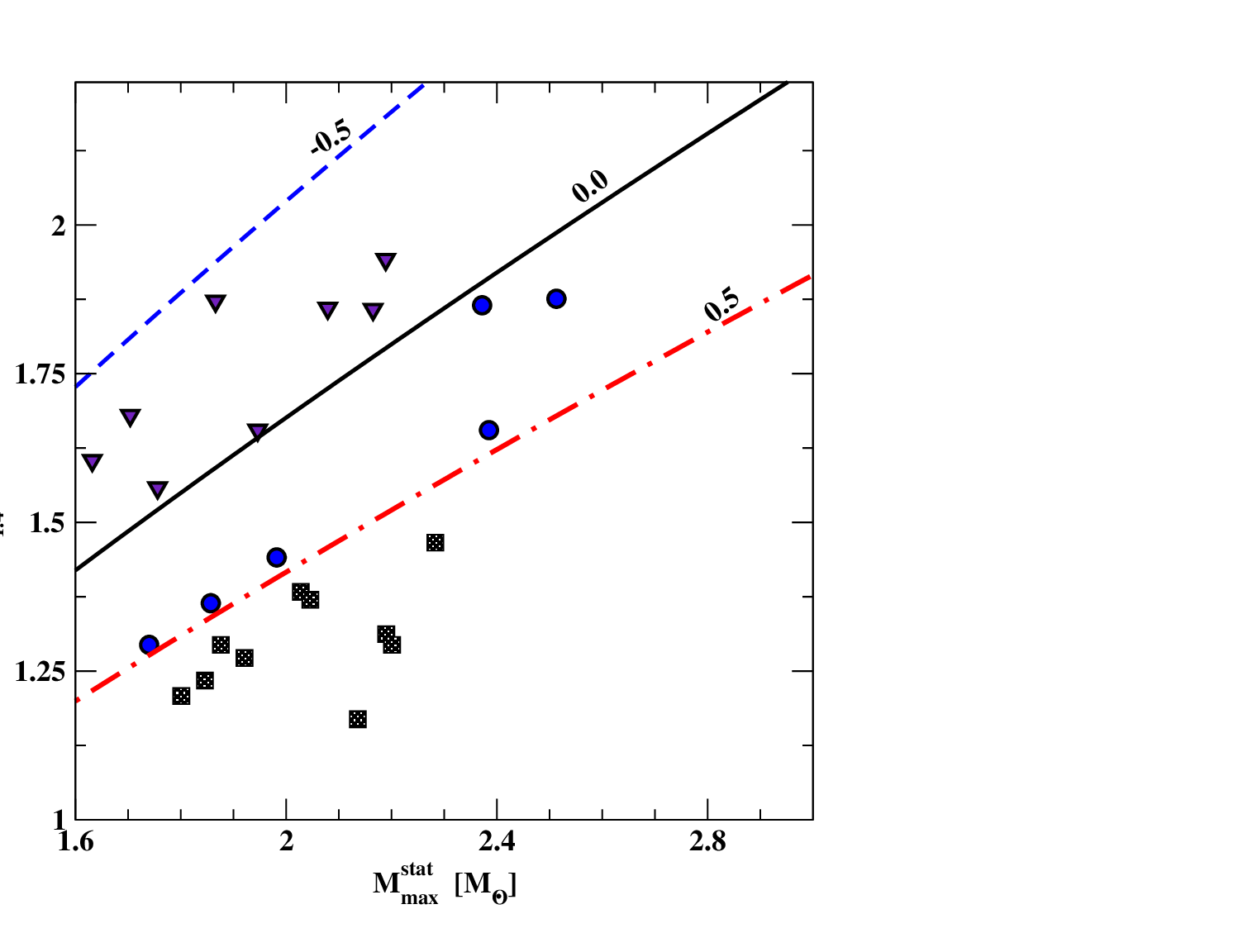}}
\caption{\label{fig:mmax_mi14} (Color online)
Same as fig. \ref{fig:mmax_r14}, but, values of moment of inertia ${\cal
I}_{1.4}^{\rm stat}$ are used instead of $R_{1.4}^{\rm stat}$.
}
   \end{figure}

We parameterize $\delta M_{\rm B}$ in terms of $M_{\rm max}^{\rm
stat}$ and $R_{1.4}^{\rm stat } $ as,
 \begin{equation}
 \frac{\delta M_{\rm B} }{M_\odot}= a_0 + a_1 \left (\frac{M_{\rm max}^{\rm
stat}}{M_\odot}\right )^\alpha
 \left (\frac{10{\rm km}}{R_{1.4}^{\rm stat}}\right )^\beta.
 \label{eq:corr}
 \end{equation}
The best fit values of the parameters appearing in Eq. \ref{eq:corr} are
calculated using the results displayed in Fig.\ \ref{fig:full_data}. The
values of parameters are  $a_0 =-2.75$, $a_1 = 2.5$, $\alpha =
0.75$ and $\beta = 1.56$. In Fig.\ \ref{fig:mmax_r14}, we plot the
results for $R_{1.4}^{\rm stat}$ versus $M_{\rm max}^{\rm stat}$
obtained by solving Eq.  \ref{eq:corr} for fixed values of $\delta M_{\rm B}$.
These plots can provide us immediately some idea of $\delta M_{\rm B}$
once the  properties of the static CS  like $M_{\rm max}^{\rm stat}$ and
$R_{1.4}^{\rm stat}$ are known.  We also superpose the results  shown
in  Fig.\ \ref{fig:full_data} by dividing them in to three classes
depending on the values of the  $\delta M_{\rm B}$.  The symbols, 
triangles, circles and squares represent the values of  $R_{1.4}^{\rm stat}$
and $M_{\rm max}^{\rm stat}$  with $\delta M_{\rm B}$ lie in the range
of $-0.5 - 0.0$, $0.0 - 0.5$ and 0.5 - 1.5 $M_\odot$ respectively. It
is evident from Fig. \ref{fig:mmax_r14} that Eqs. \ref{eq:corr} can be
used to divide the $M_{\rm max}^{\rm stat}$ $-$ $R_{1.4}^{\rm stat}$
plane in to the regions with different $\delta M_{\rm B}$.   
It is to be
noted
from Fig. \ref{fig:mmax_r14}
that the current measurement of maximum gravitational mass
1.76M$_{\odot}$ \cite{Verbiest08} would set the upper limit on
R$_{1.4}^{\rm stat}$ to be  12.4 km
which corresponds to $\delta M_B = 0$.  Interestingly, this upper limit on
R$_{1.4}^{\rm stat}$ is closer to 
the lower limit of 12.9 km obtained by analyzing the high quality X-ray spectra
from CS in qLMXB X7 \cite{Heinke06}. 
We plot in Fig.  \ref{fig:mmax_mi14} the curves for moment of inertia
${\cal I}_{1.4}^{\rm stat}$ versus $M_{\rm max}^{\rm stat}$ with fixed values of
$\delta M_{\rm B}$. These curves are generated by fitting the values of
$\delta M_{\rm B}$ to the following expression,
 \begin{equation}
\frac{\delta M_{\rm B} }{M_\odot}= a_0^\prime + a_1^\prime \left (\frac
{M_{\rm max}^{\rm stat}}{M_\odot}\right )^{\alpha^\prime}
 \left (\frac{{\cal I}_{1.4}^{\rm stat}}{{\cal I}_{0}}\right )^{\beta^\prime},
 \label{eq:corr1} \end{equation}
where, ${\cal I}_{0} = 10^{45}$g cm$^{-2}$ and the values of the best fit
parameters are  $a_0^\prime =-3.25$, $a_1^\prime = 3.25$, $\alpha^\prime =
0.63$ and $\beta^\prime = 0.85$. 
Similar to the case of R$_{1.4}^{\rm stat}$, we obtained the upper limit of ${\cal I}_{1.4}^{\rm
stat} = 1.53 \times 10^{45}$ g cm$^{-2}$ from maximum mass of CS measured
to date
\cite{Verbiest08}.

\section{Summary}

The  key properties such as $M_{\rm max}^{\rm stat}$ and $R_{1.4}^{\rm
stat}$ of static CS and $\delta M_{\rm B}$ (Eq. \ref{eq:dmb}) for the
CS rotating with the highest observed frequency of 1122Hz are computed
using 24 diverse EOSs.  These EOSs are chosen in a manner that they  correspond
to a wide variety of  approaches and their chemical composition vary from
the nucleons to hyperons and quarks in $\beta$-equilibrium.  The values
of $\delta M_{\rm B}$ are found to be almost linearly correlated with
$\left (\frac{M_{\rm max}^{\rm stat}}{M_\odot} \right)^{1/2} \left
(\frac{10{\rm km }}{R_{1.4}^{\rm stat}} \right)^{3/2}$; a combination of
$M_{\rm max}^{\rm stat}$ and $R_{\rm 1.4}^{\rm stat}$ analogous to the
one popularly used to determine Keplerian frequency.  For a given EOS,
the CS rotating at 1122Hz is non-supramassive (i.e., $\delta M_{\rm B}
> 0$) only if $\left (\frac{M_{\rm max}^{\rm stat}}{M_\odot}\right
)^{1/2} \left (\frac{10{\rm km }}{R_{1.4}^{\rm stat}} \right)^{3/2}
$ is greater than unity.  It is also noticed that the  variations in
the gravitational mass for the CS rotating with 1122Hz are up to some
extent correlated with the values of $\delta M_{\rm B}$ (see Table
\ref{tab:pro_1122}). 
In view of these results,
it appears that the observation of the rapidly rotating CSs 
constrain relative behaviour of EOS at high density with respect to it's
behaviour at low or moderate densities. Since, the $M_{\rm {max}}^{\rm stat}$
probes densest segment of the EOS, whereas, $R_{1.4}^{\rm stat}$ 
 probes relatively lower density region of EOS.  
The suitable parametric forms for the $\delta M_{\rm
B}$ in terms of $M_{\rm max}^{\rm stat}$ and $R_{1.4}^{\rm stat}$ or
${\cal I}_{1.4}^{\rm stat}$ (Eqs. \ref{eq:corr} and \ref{eq:corr1})
are also presented.  Using these parametric forms, one can divide the
$M_{\rm max}^{\rm stat}$ $-$ $R_{1.4}^{\rm stat}$ and $M_{\rm max}^{\rm
stat}$ $-$ ${\cal I}_{1.4}^{\rm stat}$ planes into regions of different
$\delta M_{\rm B}$.  Thus, for a given EOS,  only the knowledge of the
key properties of static CSs can well estimate a priori the properties
of the resulting CS rotating with 1122Hz.  
Currently measured maximum gravitational mass
1.76$M_\odot$ of PSR J0437-4715 suggests that the CS rotating at 1122Hz can
be 
non-supramassive provided
$R_{1.4}^{\rm stat} \leqslant 12.4$ km or equivalently ${\cal I}_{\rm
1.4}^{\rm stat} \leqslant
1.53 \times 10^{45}$ g cm$^{-2}$.
It will
be worth while to repeat the present investigations for the CS rotating
at higher frequencies.

\begin{acknowledgments}  The authors greatly acknowledge Professors F. Sammarruca
and J. A. Pons for providing the data for EOSs, DBHF and GMU110
respectivelly.  This work
was supported in part by the University Grant Commission under grant \#
F.17-40/98 (SA-I).  \end{acknowledgments} %

\newpage
%\bibliography{1review}

\newpage
\begin{table}
\caption{\label{tab:pro_1122}
The properties of the compact stars, rotating with frequency $\nu$
= 1122 Hz, calculated at the maximum and the minimum circumferential
equatorial radius $R_{\rm eq}^{\rm max}$ and $R_{\rm eq}^{\rm min}$.
The quantities $\delta M$ and  $\delta M_B$ for a given EOS  are
determined by using Eqs.(\ref{eq:dmb} and \ref{eq:dm}) respectively. The
central mass densities $\epsilon_c$ at the $R_{eq}^{max}$ and $R_{\rm
eq}^{\rm min}$ are also presented.}
 \begin{ruledtabular}
 \begin{tabular}{|c|ccc|ccc|cc|}
\hline EOS &$\epsilon_c$ &$M(R^{\rm max}_{\rm eq})$& $R^{\rm
max}_{\rm eq}$ & $\epsilon_c$ &$M(R^{\rm min}_{\rm eq})$&$R^{\rm min}_{\rm
eq}$ &$\delta M$ &$\delta M_B$\\

\hline
& $(10^{15}g cm^{-3})$ & $(M_\odot)$ &  (km) 
& $(10^{15}g cm^{-3})$ & $(M_\odot)$ &  (km) &$(M_\odot)$  &$(M_\odot)$ \\
\hline
GMU110 & 1.30322 &2.071 &17.53 &
  1.48038  &2.053 &16.33&0.018
      &-0.393\\
GM1-H&  1.25875 &2.235 &18.03 &
   1.56496& 2.158& 15.23&0.074
 &-0.328\\
BalbN1H1 & 1.42497& 1.825& 16.88 &
3.56904 &1.704 &9.97&0.123
&-0.079\\
G2& 1.05609& 2.051 & 17.48&
 1.81037 &2.116 &13.02 &0.065
&-0.064\\
GM1& 0.86585& 2.456 &18.55 &
 1.53015 &2.576& 13.77  &0.120
&0.007\\
UQM52&  1.14801& 1.585 &16.14&
 2.33042& 1.864 &11.48  &0.279
&0.246 \\
BJ-C& 1.05732 &1.695 & 16.46&
 2.55674 &1.954 &10.88& 0.259&
  0.278\\
GSK1&0.98210& 1.788 &16.70
& 2.37515& 2.089 &11.15 &0.301
&0.335\\
O& 0.78443 &2.138 &17.81
& 1.69183&2.554& 12.72  &0.416
& 0.461\\
SSK&  0.94170& 1.684 &16.39&
  2.55229& 2.127 &10.72   &0.443
   &0.537\\
FPS& 0.98885& 1.395& 15.51
&  3.04073& 1.881& 9.89&  0.486
   &0.583\\
BBB2&0.9615&1.476&15.79
& 2.87618& 2.002&10.08&0.526
&0.654\\
DBHF& 0.7813 &1.732 &16.54&
2.19589 &2.412 &11.24&0.680
&0.861\\
CFL52& 0.948 &0.846 &12.99
& 2.39711&1.990& 10.39 & 1.144&
 1.357\\
AU&  0.9237& 1.143&  14.46&
2.87473& 2.215& 9.76 &1.072 &1.402\\

\hline
\end{tabular}
\end{ruledtabular}
%\end{table*}
\end{table}
%__________________________________________________________________

\newpage

\newpage

\newpage

\newpage

\end{document}